\documentclass[epj-spec]{svjour}
\usepackage{graphicx,calc}
\begin{document}
\title{A continuum model for the flow of thin liquid films over intermittently chemically patterned surfaces}
\author{J.E. Sprittles\thanks{\email{sprittlj@maths.bham.ac.uk}} \and
Y.D. Shikhmurzaev\thanks{\email{yulii@for.mat.bham.ac.uk} }}
\institute{School of Mathematics, University of Birmingham,
Birmingham,  B15 2TT, UK.}
\abstract{It is known from both experiments and molecular dynamics
simulations that chemically patterning a solid surface has an effect
on the flow of an adjacent liquid.  This fact is in stark contrast
with predictions of classical fluid mechanics where the no-slip
boundary condition is insensitive to the chemistry of the solid
substrate.  It has been shown that the influence on the flow caused
by a steep change in the wettability of the solid substrate can be
described in the framework of continuum mechanics using the
interface formation theory.  The present work extends this study to
the case of intermittent patterning. Results show that variations in
wettability of the substrate can significantly affect the flow,
especially  of thin films, which may
have applications to the design of microfluidic devices.} 
\maketitle
\section{Introduction}\label{intro}
Flows of liquids on small length scales, characterised by high
surface-to-volume ratios, are of increasing interest in many
emerging technologies \cite{tabeling05}. The correct description of
the physics at liquid-solid interfaces is imperative to successful
modelling of such phenomena \cite{lauga05}.  Particular interest
lies in how modification of the chemical properties of the solid
substrate, most importantly variation in its wettability, may affect
the flow of an adjacent fluid. In classical fluid mechanics, the
no-slip boundary condition is applied on a solid substrate
irrespective of its chemical properties, whereas molecular dynamics
simulations \cite{thompson97} and experiments \cite{cottin05} have
shown that deviations from this condition can be significant, even
for chemically homogenous substrates, when considering flows on
small length scales. The conventional generalisation of no-slip is
the Navier slip condition \cite{navier23}, which states that slip (a
non-zero difference between the tangential velocities of the fluid
and substrate) is proportional to the shear stress acting on the
liquid-solid interface from the fluid.  Using the coefficient of
proportionality as an adjustable parameter, one can describe flow
over chemically homogenous solid substrates in situations where slip
is important (e.g., \cite{sokhan01}). However, the Navier condition
is a dynamic boundary condition and as such it is unrelated to the
solid's wettability, which can be defined in the absence of motion,
for example in terms of the static contact angle between a free
surface and the solid substrate.  Consequently, the Navier condition
offers no framework that would allow one to incorporate the solid's
wettability and, in this sense, has no advantages over the no-slip
condition.

It has been shown \cite{sprittles07} that the interface formation
theory \cite{shikhmurzaev07} that has previously been applied to
describe dynamic wetting and some other phenomena
\cite{shikhmurzaev05,shikhmurzaev06,blake02} can be used, without
any alterations, to study the effect of a variation in wettability
of the solid on the flow. In the present work, we consider a
particular application of the theory in the situation where the
pattern on the solid substrate consists of an array of stripes of
differing wettabilities.

\section{Problem formulation}\label{prob}
\begin{figure}
\centering
\resizebox{0.6\columnwidth}{!}{%
\includegraphics{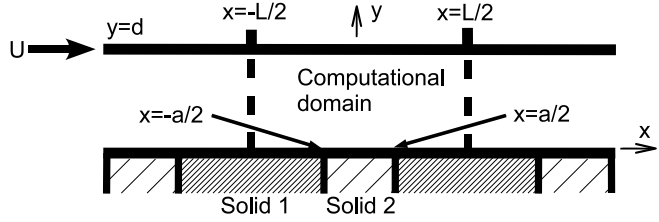} }
\caption{Problem setup.}\label{F:setup}
\end{figure}

Consider the steady two-dimensional flow of an incompressible
Newtonian liquid between two solid surfaces (Fig.~\ref{F:setup})
where the upper surface is chemically homogenous and moves parallel
to itself with velocity $\mathbf{u}=(u,v)=(U,0)$, whilst the lower
surface is chemically patterned and remains stationary. The pattern
on the lower surface consists of intermittent stripes of solid $1$,
with width $L-a$, and solid $2$, with width $a$.  It is assumed that
both surfaces are perfectly smooth. The wettabilities of solid $1$
and solid $2$ are characterised by the contact angles, $\theta_{i}$
$(i=1,2)$, which a liquid-gas free surface would form with solid
$i$.

In the bulk the flow parameters satisfy the Navier-Stokes equations,
\begin{equation}\label{ns} \nabla\cdot\mathbf{u}=0,\qquad \rho \mathbf{u}\cdot\nabla\mathbf{u} =-\nabla p
+\mu\nabla^{2}\mathbf{u},
\end{equation}
where $p$ is the pressure, $\rho$ is the density and $\mu$ is the
viscosity of the fluid.  On both surfaces we apply the boundary
conditions provided by the interface formation theory
\cite{shikhmurzaev07}:
\begin{equation}\label{state}
\sigma=\gamma(\rho^{s}_{(0)}-\rho^{s}),
\end{equation}
\begin{equation}\label{gen nav}
\mu\left(\frac{\partial u}{\partial y}+\frac{\partial v}{\partial
x}\right) +
\frac{1}{2}\left(\frac{d\sigma}{dx}+\frac{\gamma\rho^{s}
}{\rho^{s}_{e}}\frac{d\rho^{s}_{e}}{dx}\right)=\beta u, \quad
\rho v = \frac{\rho^{s}-\rho^{s}_{e}}{\tau},
\end{equation}
\begin{equation}\label{cont dens}
\frac{d\left(\rho^{s}v^{s}\right)}{dx}=
-\frac{\rho^{s}-\rho^{s}_{e}}{\tau},
\end{equation}
\begin{equation}\label{chan flow}
v^{s}=\hbox{$\frac{1}{2}$}u+\alpha\left(\frac{d\sigma}{dx}+\frac{\gamma\rho^{s}
}{\rho^{s}_{e}}\frac{d\rho^{s}_{e}}{dx}\right).
\end{equation}
A detailed explanation of the process of interface formation
together with a derivation of the equations and applications of the
theory are given in several works (e.g.
\cite{shikhmurzaev07,shikhmurzaev05,shikhmurzaev06}) so that here we
will briefly recapitulate only the important points.

At constant temperature the state of the liquid-solid interface,
i.e.\ physically, a microscopic layer of \emph{liquid} adjacent to
the solid surface, is characterised by the surface density
$\rho^{s}$ of liquid in this layer. In equilibrium one has
$\rho^{s}=\rho^{s}_{e}$ and the surface tension
$\sigma(\rho^{s}_{e})=\sigma_{e}$, where the equilibrium values
$\rho^{s}_{e}$ and $\sigma_{e}$ are different for solids of
different wettability.  The simplest equation of state accounting
for the influence of $\rho^{s}$ on $\sigma$ is given by
(\ref{state}). Instead of no-slip one now has (\ref{gen nav}), where
the first equation is the generalised Navier condition given here
for the case of a variable $\rho^s_e$ and the second describes the
mass exchange between the bulk and the liquid-solid interface. This
mass exchange is also accounted for in the mass balance equation for
the interface (\ref{cont dens}). Finally, equation (\ref{chan flow})
describes how the velocity in the interface $v^s$ depends on the
bulk velocity $u$ evaluated on the liquid-facing side of the
liquid-solid interface and the surface tension gradient. The
parameters $\alpha$ and $\beta$ characterise the response of the
interface to surface tension gradients and external torque,
respectively; $\gamma$ is associated with the inverse
compressibility of the fluid; $\rho^s_{(0)}$ is the surface density
corresponding to zero surface tension; $\tau$ is the surface tension
relaxation time; in the simplest variant of the theory these are all
treated as material constants.

The terms including $d\rho^s_e/dx$ in (\ref{gen nav}) and (\ref{chan
flow}) describe the surface force which ensures that, in the case of
a variable $\rho^s_e$ (and hence $\sigma_e$), there is no perpetual
motion. Once the outer flow drives the matter in the liquid-solid
interface along a solid surface whose wettability varies, the
surface tension gradient in (\ref{gen nav}) and (\ref{chan flow})
becomes unbalanced triggering the interface formation process which
in its turn affects the flow that caused it.

The last step now is to specify the wettability pattern of the lower
surface, i.e.\ prescribe the dependence of $\rho^s_e$ on coordinates
along it and the value of $\rho^s_e$ on the upper surface . From the
viewpoint of fluid mechanics, the junction between solid $1$ and
solid $2$ is associated with a discontinuity in the properties of
the substrate and, therefore, requires special handling. In order to
avoid this, we consider a transition region in which the equilibrium
surface density varies smoothly, from its value $\rho^{s}_{1e}$ on
solid $1$ to value $\rho^{s}_{2e}$ on solid $2$ (or vice-versa),
across a finite distance $l$. This can be modelled, for example,
using
\begin{equation}\label{rhose} \rho^{s}_{e}=\rho^{s}_{1e}+
\hbox{$\frac{1}{2}$}\left(\rho^{s}_{2e}-\rho^{s}_{1e}\right)
\left\{\tanh[(a/2+x)/l] +\tanh[(a/2-x)/l]\right\}.
\end{equation}
On the upper solid we assume $\rho^{s}_{e}\equiv\rho^{s}_{1e}$.

Using the Young equation,
$\sigma(\rho^s_{ie})=-\sigma_{lg}\cos\theta_{i}$, where
$\sigma_{lg}$ is the surface tension of the liquid-gas free surface,
and the equation of state (\ref{state}) one can relate $\rho^s_{1e}$
(or $\rho^s_{2e}$) with the equilibrium contact angle $\theta_1$ (or
$\theta_2$) which the free surface would form with solid 1 (or 2):
\begin{equation}\label{rhotheta}
\rho^{s}_{ie}=\rho^{s}_{(0)}
+\gamma^{-1}\sigma_{lg}\cos\theta_i,\qquad(i=1,2).
\end{equation}
The contact angle is the conventional and convenient measure of the
solid surface wettability with respect to a given fluid, and in what
follows we use $\theta_{i}$ instead of $\rho^{s}_{ie}$.

Due to the periodicity of the problem it is convenient to use
periodic boundary conditions
\begin{equation}
\label{periodic}
\mathbf{u}\mid_{x=-L/2}=\mathbf{u}\mid_{x=L/2},~~~
\frac{\partial\mathbf{u}}{\partial
x}\mid_{x=-L/2}=\frac{\partial\mathbf{u}}{\partial x}\mid_{x=L/2},
\end{equation}
\begin{equation}\label{sym_nd2}
\rho^{s}\mid_{x=-L/2}=\rho^{s}\mid_{x=L/2},~~~
v^{s}\mid_{x=-L/2}=v^{s}\mid_{x=L/2}.
\end{equation}

Equations (\ref{ns})--(\ref{rhose}),
(\ref{periodic})--(\ref{sym_nd2}) now fully specify our problem.

\section{Results}
The problem has been solved numerically using the finite element
method. After using $L$, $d$, $U$, $dU/L$, $\mu UL/d^{2}$,
$\sigma_{lg}$ and $\rho^{s}_{(0)}$ as scales for the horizontal
coordinate, vertical coordinate, tangential velocities, vertical
velocity, pressure, surface tension and surface density,
respectively, one has that the problem is specified by following
non-dimensional parameters:
$$
r=\frac{d}{L},\ \bar{a}=\frac{a}{L},\ \ Re=\frac{\rho U d}{\mu},\
Ca=\frac{\mu U}{\sigma_{lg}},\ \epsilon=\frac{U\tau}{d},\
\bar{\beta}=\frac{\beta U d}{\sigma_{lg}},\
$$
$$
Q=\frac{\rho_{(0)}^{s}}{\rho U\tau},\ \bar{\alpha}=\frac{\alpha
\sigma_{lg}}{U d},\ \bar{l}=\frac{l}{L},\
\lambda=\frac{\gamma\rho^{s}_{(0)}}{\sigma_{lg}},\
\theta_{i}\quad(i=1,2).
$$

As the interface formation theory is applied here without any
\emph{ad hoc} alterations, we may take estimates for the theory's
phenomenological constants from independent experiments. Dynamic
wetting experiments \cite{blake02}, for example, suggest that for a
fluid with $\mu\sim10$ g~cm$^{-1}$~s$^{-1}$ and $\rho\sim 1$
g~cm$^{-3}$ we have $\alpha\sim10^{-8}$ g$^{-1}$~cm$^{2}$~s,
$\beta\sim 10^{8}$ g~cm$^{-2}$~s$^{-1}$, $\rho^{s}_{(0)}\sim
10^{-7}$ g~cm$^{-2}$ and $\tau\sim 10^{-7}$ s.  Using these
estimates and taking typical values associated with the flows of
thin liquid films, $U\sim 10$ cm s$^{-1}$, $d\sim 10^{-5}$ cm,
$\sigma_{lg} \sim 10^{2}$ dyn~cm$^{-1}$ and $l\sim 5\times10^{-7}$
cm, we have the following typical magnitudes of the non-dimensional
groups: $Re=10^{-5}$, $Ca=1$, $\epsilon=10^{-1}$,
$\bar{\beta}=10^{2}$, $Q=10^{-1}$, \ $\bar{\alpha}=10^{-2}$,
$\bar{l}=5\times 10^{-2}$ and $\lambda=2$.

\begin{figure}\centering
  \begin{minipage}{0.7\textwidth}
     \includegraphics[width=\textwidth]{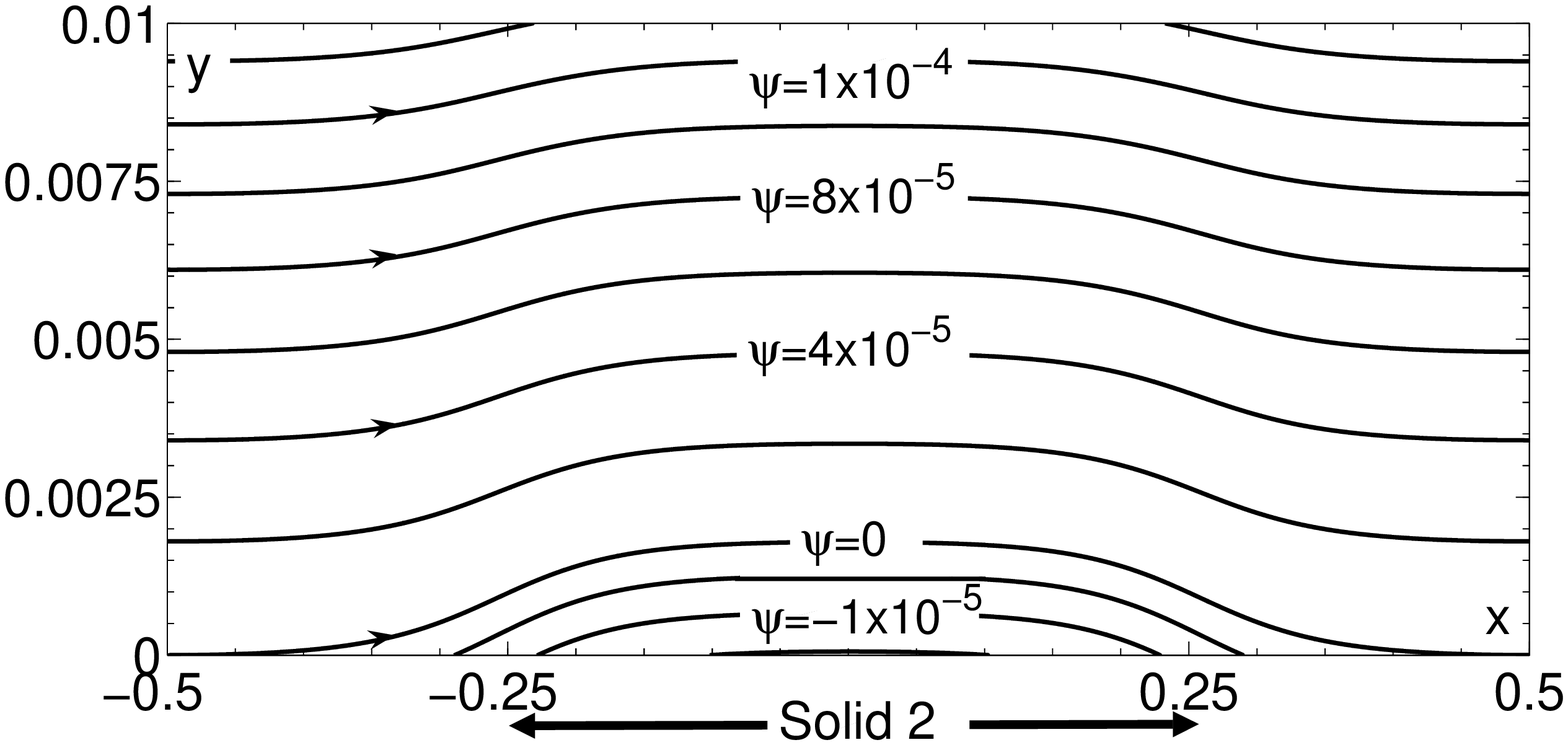}
  \end{minipage}
  \begin{minipage}{0.27\textwidth}
\includegraphics[width=\textwidth]{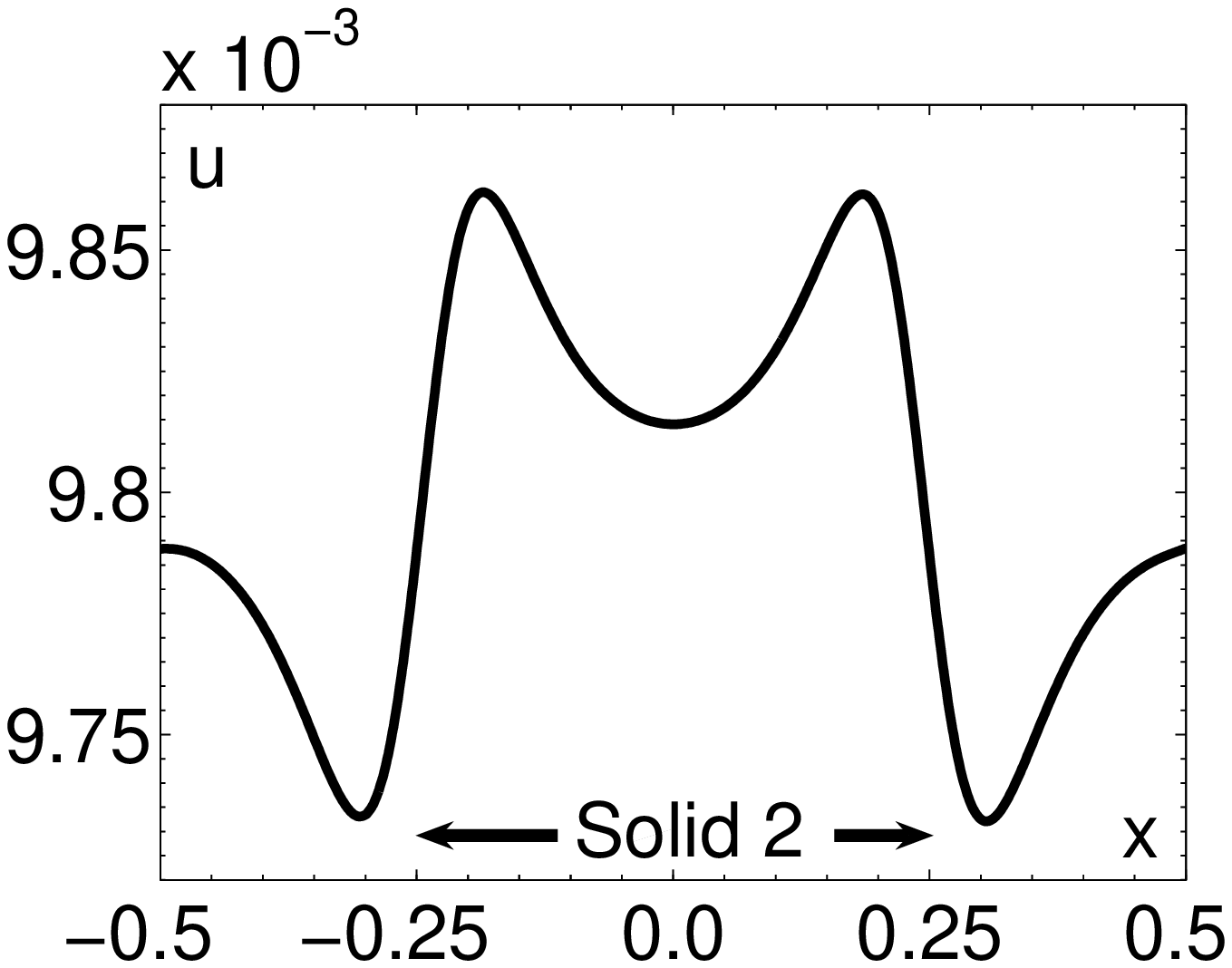}
\\
\includegraphics[width=\textwidth]{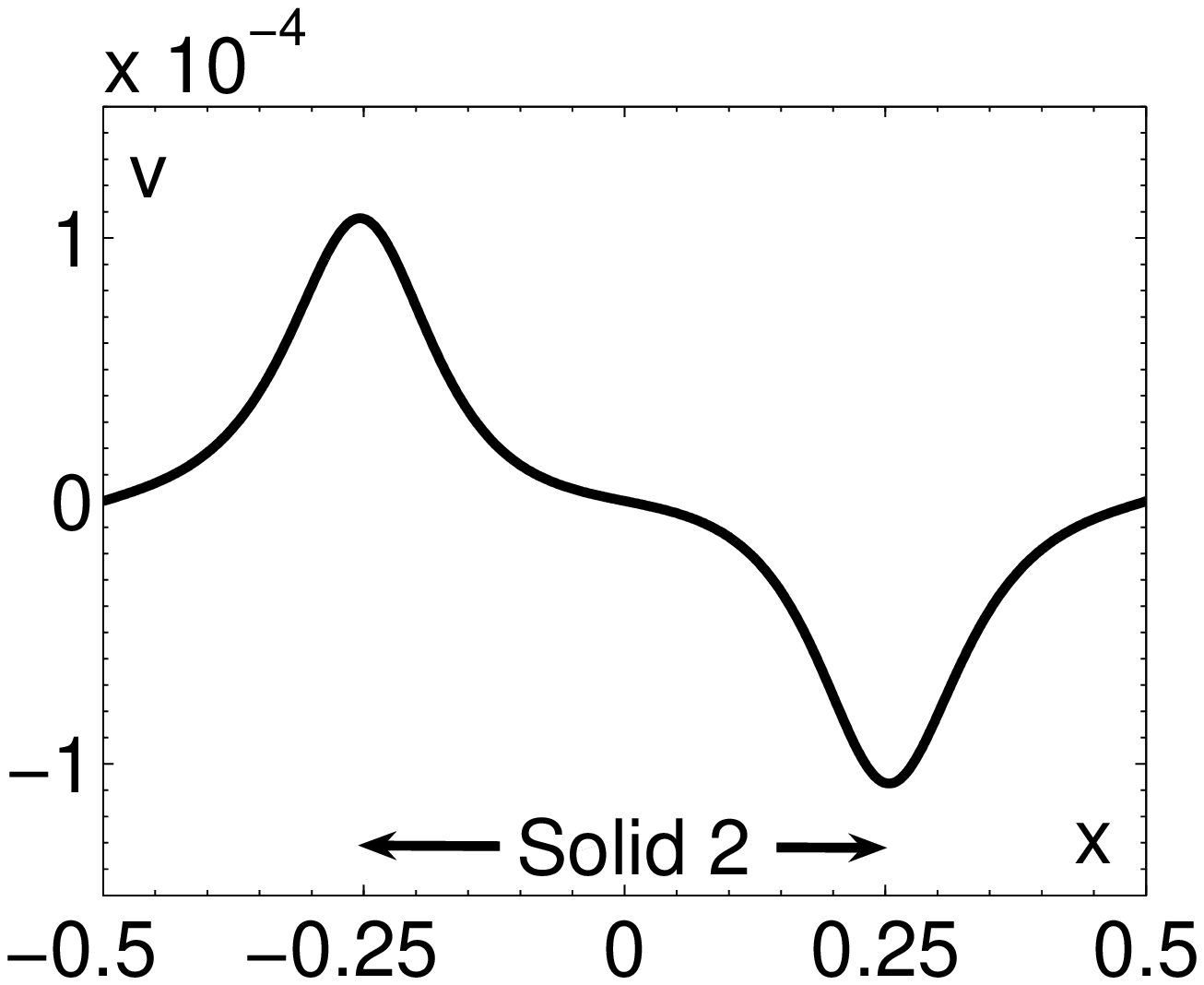}
 \end{minipage}
\caption{Streamlines and the corresponding components of velocity on
the substrate for flow over a intermittently chemically patterned
surface in which values of the streamfunction $\psi$ are given.
Parameters are at their typical values with $\theta_{1}=10^{\circ}$
and $\theta_{2}=80^{\circ}$.}\label{F:pic}
\end{figure}

Consider each stripe of solid to have equal width, $a=1/2$, and the
aspect ratio $r=1$.  Then, without loss of generality, we may
consider solid $1$ to be more hydrophilic than solid $2$.
Streamlines for the typical values and $\theta_{1}=10^{\circ}$,
$\theta_{2}=80^{\circ}$ are shown in Fig.~\ref{F:pic}. The less
hydrophilic substrate appears to acts as an obstacle to motion with
streamlines coming out of the surface as the solid becomes less
hydrophilic, at $x=-0.25$, and entering again as the solid becomes
more hydrophilic again, at $x=0.25$. This effect can be seen clearly
by looking at the normal velocity on the liquid facing side of the
liquid-solid interface in Fig.~\ref{F:pic}. One can see that the
components of velocity are trying to relax to their equilibrium
values, $\mathbf{u}_{e}=(9.8\times 10^{-3},0)$ but are unable to
obtain them before the next transition in wettability. This means
that the distance between transitions is of the same order as the
relaxation length, which is the distance that it takes for surface
properties to relax to their equilibrium values. The width of the
transition region $\bar{l}$ is far smaller than the relaxation
length and, as in \cite{sprittles07}, its exact size has very little
effect on the overall flow field.

Notably, in contrast to studies of slip on super hydrophobic
substrates, where the presence of nanobubbles leads to the effective
slip length of the system becoming a relevant measure of the effects
of chemically patterning the substrate \cite{lauga03,priezjev05},
here the slip coefficient $\beta$ remains the same on both solids.
Importantly, as shown in \cite{sprittles07}, slip results primarily
from the disturbance of the force balance in the liquid-solid
interface and not from the tangential stress, as follows from the
standard Navier condition. This has significant consequences when
attempting to interpret the results of molecular dynamics
simulations and experiments in terms of a continuum theory.

\emph{Acknowledgement.}  The authors kindly acknowledge the
financial support of Kodak European Research and the EPSRC via a
Mathematics CASE award.

\end{document}